\documentclass[preprint,showpacs,preprintnumbers,amsmath,amssymb]{revtex4}

\usepackage{graphicx} 
\usepackage{dcolumn}  
\usepackage{bm}       

\newcommand{\la}{\langle}
\newcommand{\ra}{\rangle}

\begin{document}
\preprint{}   

\title{Transport Coefficients of Gluon Plasma}

\author{Atsushi Nakamura}
 \email{nakamura@riise.hiroshima-u.ac.jp}
 \affiliation{%
RIISE, Hiroshima University, Higashi-Hiroshima, 739-8521, Japan
and \\
Institute for Nuclear Theory, University of Washington, Box 351550, Seattle
WA 98195, USA} 
\author{Sunao Sakai}
 \email{sakai@e.yamagata-u.ac.jp}
 \affiliation{%
Faculty of Education, Yamagata University, Yamagata 990-8560, Japan
}


\begin{abstract}
Transport coefficients of gluon plasma are calculated for
a SU(3) pure gauge model by lattice QCD simulations 
on $16^{3} \times 8$ and $24^{3} \times 8$ lattices. 
Simulations are carried out at a slightly above the deconfinement 
transition temperature $T_{c}$, 
where 
a new state of matter is currently being pursued in RHIC experiments.
Our results show that the ratio of the shear viscosity to the entropy
is less than one
and the bulk viscosity is
consistent with zero in the region, $ 1.4 \leq T/T_c \leq 1.8 $. 
\end{abstract}


\maketitle

\noindent
{\bf Introduction:}
RHIC experiments have been bringing us many surprises. 
One of them is that the data are unexpectedly well described 
by the hydrodynamical model \cite{Star01}. 
Experimental data and phenomenological analyses suggest 
the quark-gluon plasma (QGP), or a new
state of matter may be produced.
See Ref.\cite{Ludlam} for a review of RHIC experiments.
Molnar and Gyulassy investigated the elliptic flow data using 
a Boltzmann-type equation for gluon scattering, and found that they
needed a cross section about 50 times larger than expected in the
perturbative QCD\cite{Molnar02}.  
This indicates that the QGP state above the
phase transition temperature, $T_c$, is not a free gas of
perturbed gluons.
QCD-TARO collaboration measured the temporal meson propagators
and found that their wave functions do not behave as free
particles even at $T\sim1.5T_c$; they conjectured 
that the strong interactions between the thermal gluons and quarks may
provide binding forces.
Recently, more extended analyses of the temporal propagators were
reported by three groups \cite{Hatsuda03,Datta03,Umeda02} and it was 
suggested that the charmonium state survives until around  $2T_c$.

The new state of matter produced at high temperatures in RHIC experiments
is most likely not a weakly interacting plasma, 
but a strongly interacting quark-gluon system. 
Investigating the results in Ref.\cite{Molnar02}, Teaney 
found that $\eta/s \sim 0.04$,
where $\eta$ and $s$ are the shear viscosity and the entropy,
respectively\cite{Teaney03}.
Shuryak and Zahed have proposed a `strongly coupled QGP' model for 
the new state of matter above $T_c$\cite{ShuryakZahed}, and
argued that the QGP studied in RHIC is the most perfect fluid
ever measured.
Policastro et al. have calculated $\eta$ for
the finite-temperature ${\cal N}=4$ supersymmetric Yang-Mills theory in 
the large $N$, strong-coupling regime, and obtained 
$\eta/s=1/4\pi$\cite{PTS01}.  
This value is found to be universal for theories with
gravity duals
and it is conjectured that $\eta/s=1/4\pi$
is a lower limit for all systems in nature\cite{KSS04}.

It has now become highly desirable to study the nature 
of the quark gluon system,
particulaly its hydrodynamical parameters such as the transport coefficients 
above $T_c$ based on QCD in a non-perturbative manner.
In this paper, we calculate the transport coefficients
of QGP at a slightly above $T_{c}$, from the lattice simulations.
Simulations are carried out in the quench approximation.
For the calculation of the transport coefficients on a lattice, 
we apply the formulation based on the linear response
theory\cite{Horsley,Zubarev,Hosoya},
where the transport coefficients are calculated from
Matsubara Green's function of energy momentum tensors. 
Numerical simulations of transport coefficients with this formulation
were first carried out by Karsch and Wyld\cite{Karsch}.
In their pioneering work, they performed the simulation on an $8^3 \times 4$ 
lattice but unfortunately, the size in the imaginary time direction 
was too small for the determination of the transport coefficients. 

We report here our simulation on a $N_{T}=8$ lattice with
RG (renormalization group) improved action by Iwasaki.
Our results are summarized as follows.
\begin{enumerate}
\item
The ratio of the shear viscosity to the entropy, 
$\eta/s$, is small, 
i.e., less than one, but it is most probably larger than $1/4\pi$.
See Fig.\ref{EtaS}.
\item
The bulk viscosity is less than the shear viscosity and
is consistent with zero within the present statistics.
\item
For the heat conductivity, we could obtain no meaningful
result.
This is because, in pure gauge theory, there is no
conserved current which transports the heat.  
\footnote{We thank Dam Son for pointing out this fact.}
\end{enumerate}
Preliminary results based on $16^3\times 8$ and smaller lattices 
have been  reported at lattice and Quark Matter conferences\cite{QM97,sakaist}.

\bigskip
\noindent
{\bf Transport Coefficients in Linear Response Theory:}
The formulation for the transport coefficients of QGP 
in the framework of the linear response theory has been given
in Refs.\cite{Horsley,Zubarev,Hosoya}.
For the sake of consistency, we shall summarize the formula
which will be used in the following calculations.

Transport coefficients are calculated
using the space-time integral of a retarded Green's function of energy
momentum tensors,
\begin{equation}
\eta
 = - \int
<T_{12}(\vec{x},t)T_{12}(\vec{x}',t')>_{ret} ,
\label{shear}
\end{equation}
\vspace{-3mm}
\begin{equation}
\frac{4}{3}\eta+\zeta
 = -\int
<T_{11}(\vec{x},t)T_{11}(\vec{x}',t')>_{ret} ,
\label{bulk}
\end{equation}
\vspace{-3mm}
\begin{equation}
\chi
 = -\frac{1}{T} \int
<T_{01}(\vec{x},t)T_{01}(\vec{x}',t')>_{ret} ,
\label{heat} 
\end{equation}
\noindent
where 
$\int \equiv 
\int d^{3}x' \int_{-\infty}^{t} dt_{1}  e^{\epsilon(t_{1}-t)} 
\int_{-\infty}^{t_{1}}dt'$, 
and
$\eta$, $\zeta$ and $\chi$ represent shear viscosity, 
bulk viscosity and heat conductivity, respectively.
$<T_{\mu \nu} T_{\rho \sigma}>_{ret}$ is the retarded Green's
function of the energy momentum tensors at finite temperature.
For the pure gauge theory, $T_{\mu \nu}$'s are written
by the field strength tensors $F_{\mu\nu}$: 
\begin{equation}
T_{\mu\nu}= 2 \,\mbox{Tr}\, [F_{\mu\sigma}F_{\nu\sigma}
-\frac{1}{4}\delta_{\mu\nu}F_{\rho\sigma}F_{\rho\sigma}]\label{tensor}.
\end{equation}
\noindent
$F_{\mu \nu}$ are defined by plaquette 
variables on the lattice as
$ U_{\mu\nu}(x) =  \exp{(ia^2 g F_{\mu \nu}(x))} $.
$F_{\mu \nu}$ are obtained either by taking the $log$ of $U_{\mu \nu}$
directly, or by expanding $U_{\mu \nu}$ with respect to $a^{2} g$.  
In the following, we use the latter method to calculate 
$F_{\mu\nu}$\cite{Karsch}.
\footnote{In studying the $SU(2)$ case,
we have observed little difference in Matsubara Green's function 
between the two definitions of $F_{\mu \nu}$\cite{sakaist}.}

It is difficult to calculate the retarded
Green's functions in the lattice QCD, in which
Matsubara Green's functions are measured.
The retarded Green's functions are obtained by the analytic continuation. 
We obtain the numerical values of Matsubara Green's functions at
discrete variables $\omega_{n}=2\pi nT$ in the momentum space, 
while the retarded Green's  functions are functions of the continuous
variable $p_{0}$. Therefore, we need a bridge 
for the analytic continuation.

Matsubara Green's functions $ G_{\beta}$ are expressed in 
a Fourier transformed form
with the spectral function $\rho$:
\begin{equation}
G_{\beta}(\vec{p},t)=
\sum_n e^{i\omega_n t} 
\int d\omega
\frac{\rho(\vec{p},\omega)}{i\omega_{n}-\omega} .
\label{spector}
\end{equation} \noindent
It is well known that
the spectral function is common to both the retarded and Matsubara
Green's functions\cite{HNS93}. 
The expression for the retarded Green's functions is obtained
by putting $\omega \rightarrow p_{0}+i\epsilon$.

The determination of $\rho(\vec{p},\omega)$ is not
straightforward,
because in a numerical simulation, Matsubara Green's function has a finite
number of points in the temperature direction,  $N_{T}/2$. 
We must employ an ansatz for the spectral function with parameters,
which are determined by fitting  Matsubara Green's function. 
The simplest nontrivial ansatz for the spectral function has been 
proposed by Karsch and Wyld\cite{Karsch},
\begin{equation}
\rho(\vec{p}=0,\omega)=\frac{A}{\pi}
(\frac{\gamma}{(m-\omega)^2+\gamma^2}-
\frac{\gamma}{(m+\omega)^2+\gamma^2}),
\label{ansatz}
\end{equation}\noindent
where $\gamma$ represents the effects of interactions and is related 
to the imaginary part of the selfenergy. This ansatz is supported by
perturbative calculations\cite{Horsley,Hosoya}.

Once we use this ansatz for the spectral function, the space time
integral of the retarded Green's function can be calculated analytically.
The result is
\begin{equation}
 \alpha = 2A\frac{2\gamma m}{(\gamma^2+m^2)^2} ,
\label{TP} 
\end{equation}
where
$\alpha$ represents the shear viscosity $\eta$, bulk viscosity $\zeta$
or heat conductivity $\chi$ times $T$.
At least three independent data points for Matsubara Green's
functions are necessary to determine 
these parameters. 

In Ref.\cite{Karsch}, a simulation was carried out 
on a $8^3 \times 4$ lattice, where two independent 
data points in the temperature direction are available. 
In this simulation,
three parameters in the spectral function could not be determined. 
In order to determine $A$, $\gamma$ and $m$, we adopt
$N_{T}=8$.

\bigskip
\noindent
{\bf Numerical Simulations:}
We calculate the transport coefficients in the $SU(3)$
gauge theory for the regions a slightly above the transition temperature
which are covered in RHIC experiments.
We adopt Iwasaki's improved gauge action, 
which is closer to the renormalized trajectory than
the plaquette action, and
we obtain results close to the
continuum limit on relatively coarse lattices\cite{Yoshie}.
We found that the fluctuation of the Matsubara Green's
function is much suppressed comparing with the standard plaquette
action\cite{sakaied}. 

We should first determine the critical $\beta$ of Iwasaki's improved
action on the $N_{T}=8$ lattice.  
For the $N_{T}=4$ and $6$ lattices, the critical $\beta$ for
this action were determined by the Tsukuba group\cite{Kaneko}.
We have carried out a simulation for $\beta_{c}$ on 
a $16^{3} \times 8$ lattice; the results were
reported in Ref.\cite{sakaied}. 
However, the volume size was small,
and we could obtain only a rough estimation of $\beta_{c}$, 
that is, $2.70 < \beta_{c}< 2.72 $. 
If we use the finite size scaling formula
reported by the Tsukuba group, $\beta_{c}$ at $N_{T}=8$ becomes 
$ 2.72< \beta_{c} < 2.74$. 
The values of $\beta_{c}$ determined by
the simulation for $N_{T}=4,6,8$ do not yet satisfy the asymptotic
two-loop scaling relation.
We take $\beta=3.05$, $3.2$ and $3.3$ as our
simulation points.

\begin{figure}[htb]
\begin{center}
\includegraphics[width=0.8\linewidth]{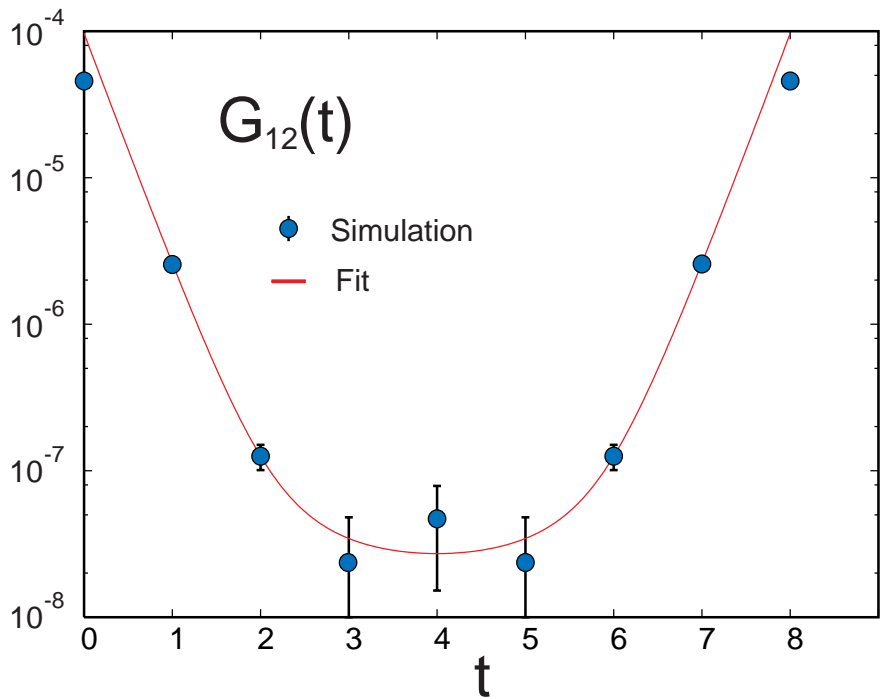}
\vspace{-5mm}
\caption{Numerical data points and fitting results of
Matsubara Green's function $G_{12}(t)$ on a $24^3 \times 8$ lattice 
\label{fig1}}
\end{center}
\vspace{-5mm}
\end{figure}

\begin{figure}[htb]
\begin{center}
\includegraphics[width=0.8\linewidth]{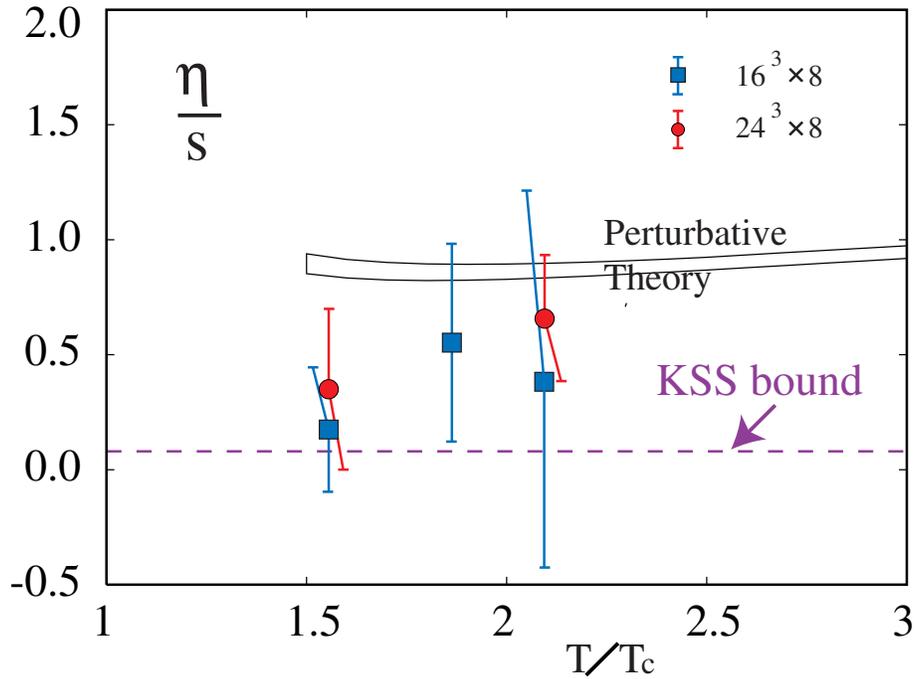} 
\vspace{-5mm}
\caption{The ratio of the shear viscosity to the entropy
as a function of $T/T_c$
KSS bound is $1/4\pi$\cite{KSS04}. `Perturbative theory' is constructed from
$\eta$ in Ref.\cite{Arnold03} and $s$ in Ref.\cite{Blaizot99}.
\label{EtaS}}
\end{center}
\end{figure}

\bigskip
\noindent
\underline{Matsubara Green's Function on $N_{T}=8$ Lattice:}
The parameters of the simulations and the obtained 
statistics are summarized in Table 1.
For Matsubara Green's functions $G_{11}$ and $G_{12}$ from which the
shear and bulk viscosities are calculated, we can obtain reliable
signals from 
approximately $0.8 \times 10^{6}$ MC data on a $24^3 \times 8$ lattice.
As an example, $G_{12}$ is shown in Fig.\ref{fig1} for $\beta=3.3$.
\begin{center}
\renewcommand{\arraystretch}{0.9}
\begin{table}[h]
\begin{tabular}{|c|c|c|c|c|c|c|c|}
     \hline
     & \multicolumn{1}{|c|}{$\beta$} &
     \multicolumn{1}{|c|}{total sweeps} &
     \multicolumn{1}{|c|}{For equilibrium} &
     \multicolumn{1}{|c|}{bin size} \\
     \hline
     \hline
       & 3.05 &1333900 &133900 &100000 \\      
$16^3$ & 3.2  &1212400 &112100 &100000 \\
       & 3.3  &1265500 &165500 &100000 \\
     \hline
$24^3$ & 3.05 & 861000 & 61000 &100000 \\
       & 3.3  & 784000 & 84000 &100000 \\
     \hline
\end{tabular}
\caption{Simulation parameters and statistics. Data at $t=0$ and $t=8$
are not used for the fit. 
\label{simpara} }       
\vspace{-0.5cm}
\end{table}
\end{center} 
In the case of the $16^{3}\times 8$ lattice, the errors are larger
than the signal at $\tau=4$, even with
more than $10^{6}$ Monte Carlo (MC) data.
The volume of $16^3$ may be too small for $N_{T}=8$.

$G_{14}$, from which the heat 
conductivity is calculated, has too large a background noise
to extract a signal.  
Therefore, the fitting of Matsubara Green's function 
by the spectral function of Eq.(\ref{ansatz}) is
carried out only for $G_{11}$ and $G_{12}$.

\bigskip
\noindent
\underline{Transport Coefficients of Gluon Plasma:}
The fitting of Matsubara Green's function by Eq.(\ref{ansatz})
was carried out by applying a non-linear least-square fitting program, SALS.
Then the transport coefficients of the gluon plasma are calculated
using Eq.(\ref{TP}). The errors are estimated by
the jackknife method.
After equilibrium is reached, the data 
are grouped into bins and the average of the data
in each bin is treated as an independent data sample. 
The bin size is changed from $5\times 10^4$ to $1.2 \times 10^5$. 
The results are independent of the bin size.
In the following, 
the bin size is $100000$ as shown in Table \ref{simpara}. 
%
%
\begin{center}
\renewcommand{\arraystretch}{0.9}
\begin{table}[h]
\begin{tabular}{|c|c|c|c|c|c|c|}
     \hline
     & \multicolumn{1}{|c|}{$\beta$} &
     \multicolumn{1}{|c|}{$\eta a^3$} &
     \multicolumn{1}{|c|}{$\zeta a^3$} &
     \multicolumn{1}{|c|}{$\eta$ $\mbox{GeV}^3$} & 
     \multicolumn{1}{|c|}{$\zeta$ $\mbox{GeV}^3$} \\
     \hline

     \hline
       & 3.05 &0.0018(28) & -0.0015(29) & 0.054(82) &-0.044(85)\\
$16^3$ & 3.2  &0.0059(46) & -0.0025(20) &0.281(223) &-0.122(90)\\
       & 3.3  &0.0043(90) &-0.0041(142) &0.283(590) &-0.027(931)\\
     \hline
$24^3$ & 3.05 &0.0036(36) &-0.00095(288)  &0.106(108)  &-0.028(85)\\
       & 3.3  &0.0072(30) &  -0.0031(26)  &0.471(194)  &-0.201(167)\\
%
      \hline
\end{tabular}
\caption{Shear and bulk viscosities 
in non-dimension and in the physical units.
The lattice scales, $a^{-1}=$ 3.09, 3.62 and 4.03 GeV 
for $\beta=$ 3.05, 3.20 and 3.30, respectively.  
\label{simtp} }       
\end{table}
\end{center} 
The results for the shear and bulk
viscosities  are given Table \ref{simtp}.
The bulk viscosity is equal to zero within error bars, 
while the shear viscosity remains finite.
We do not see the size dependence.

In the lattice calculations, the shear viscosity is calculated
in the form
$\eta \times a^3$. In order to express it in physical units,
we should know the lattice spacing $a$ at each $\beta$ value.  
For the estimation of $a$, we use the finite temperature transition point 
$\beta_c$. 
We take $\beta_{c}=2.73$ for $N_{T}=8$.
The transition temperature is $T_{c}=276 MeV$\cite{Kaneko}, and assume
asymptotic two-loop scaling for the region $\beta > 2.73$.
The lattice spacing and the shear and bulk viscosities 
in the physical units are also listed in Table \ref{simtp}.
$\eta^{1/3}$ expressed in the physical units are 
slightly less than the ordinary hadron masses around $T_{c}$.

\bigskip
\noindent
\underline{Entropy density:}
In a homogeneous system, the free energy has the form of
$F=fV$, and then the pressure is $p=-f$.
Using the thermodynamical relation,
$U-TS=-T\log Z=F$, we obtain\cite{Boyd96} 
\begin{equation}
s = S/V = (\epsilon+p)/T,
\label{Entropy}
\end{equation}
where $\epsilon$ is the energy density.
Using lattices with $N_T=8$ and the integration method,
$p/T^4|_{\beta_0}^{\beta}=\int_{\beta_0}^{\beta}d\beta'
N_T^4(\la S\ra_T-\la S\ra_0)$,
CP-PACS obtained $p$ and $\epsilon$, 
where $\la S\ra_T$ and $\la S\ra_0$ are the expectation values 
of the action density at temperature $T$\cite{CPPACS99}.
We reconstruct the results from their numerical data of
$\la S\ra_T$ and $\la S\ra_0$ and calculate the entropy density
in Eq.(\ref{Entropy}).

\bigskip
\noindent
{\bf Concluding Remarks:}
In the high temperature limit, the transport coefficients have been
calculated
analytically by the perturbation method
\cite{Kajantie,Horsley,Gavin,BMPR90,Arnold00,Arnold03}.
They are summarized as follows.\\
(1) The bulk viscosity is smaller than the shear viscosity.
This is consistent with our numerical results.
(2) The shear viscosity in the next-to-leading-log is expressed by
\cite{Arnold03},
$
\eta_{NLL} =({T^{3}}/{g^4}) {C_1}/{\log(\mu^*/m_D)}
$
where $m_D=\sqrt{1+N_f/6}gT$, and for the pure gluon system 
$C_1=27.126$ and $\mu^*/T=2.765$
.

There is a slight ambiguity in the relationship between coupling $g$
and the temperature, and we use a simple form,
$g^{-2} = 2b_0 \log(4T/\Lambda)$ with $b_0=11N_c/48\pi^2$.
The scale parameter $\Lambda$ on the lattice is 
set to be
$\Lambda/T_{c} \simeq 1.5$.
For the entropy density, we use a hard-thermal loop
result\cite{Blaizot99}.
With these formulae, the perturbative $\eta/s$ 
can be compared with the results of numerical calculations. 
The result is shown in Fig.\ref{EtaS}. 

In this letter, we report  the first lattice QCD result of the 
transport coefficients in the vicinity of the critical temperature. 
Although it still contains large errors,
 it may provide useful information for
understanding QGP in these temperature regions.
In particular, a small $\eta/s$ supports the success of the hydrodynamical
description for QGP.
Applicability conditions of the hydrodynamical model in
quantum field theory were first considered in Ref.\cite{Namiki59}.
Together with experimental and phenomenological studies, 
the field theoretical approach will enrich our understanding of the new
state of matter.
We have shown here that the lattice QCD numerical simulations
can provide useful information. 

The next step is to obtain data with smaller systematic and statistical
errors.  
If we can reduce the error bars in Fig.\ref{EtaS}
by a factor of two or three, 
we may realistically compare the data with the conjecture in 
Ref.\cite{KSS04}.
We observed that Matsubara Green's
function suffer from large fluctuations, but by using
the improved action the fluctuations are significantly reduced. 
Another possibility for reducing the fluctuations may be to employ
improved operators for $T_{\mu\nu}$\cite{Tsumura}.

The results here depend on the ansatz of the spectral function of 
the Fourier transform of Matsubara Green's function. 
In order to test the functional form of the spectral function, 
we need more data points for Matsubara Green's function 
in the temperature direction, for which 
the most effective approach will be to apply an anisotropic 
lattice.
If we have sufficient data points, the maximum entropy method 
is a promising way of determining the spectral function\cite{MEM00}
which is free from the ansatz.
Aarts and Resco pointed out, however, that it is difficult to extract
transport coefficients in weakly-coupled theories from the euclidean
lattice, since Green's function is insensitive to details of
the spectral function $\rho(\omega)$ at small $\omega$\cite{Aarts02}.
New concepts will be necessary to overcome this diffculty.

\noindent
{\bf Acknowledgement}
We thank Tetsuo Hatsuda and Kei Iida for many useful discussions 
and their constant encouragement.  
One of the authors (A.N.) would like to thank Andrei Starinets 
for his kind and patient 
explanations of Refs.\cite{PTS01} and \cite{KSS04} for the
author who is ignorant of this field. 
The simulations were carried out at KEK and at RCNP.

\end{document}